\documentclass[12pt,aps,preprint,nofootinbib]{revtex4}

\usepackage{graphicx}
\usepackage{cancel}
\usepackage{amssymb}
\usepackage{textcomp}
\usepackage{amsmath}
\usepackage{bm}
\usepackage{times}
\usepackage{epsfig}
\usepackage{epstopdf}
\usepackage{color}
\usepackage{multirow}
\usepackage[colorlinks=true, pdfstartview=FitV, linkcolor=blue, citecolor=blue, urlcolor=blue]{hyperref}

\def\lsim{\mathrel{\raise.3ex\hbox{$<$\kern-.75em\lower1ex\hbox{$\sim$}}}}
\def\gsim{\mathrel{\raise.3ex\hbox{$>$\kern-.75em\lower1ex\hbox{$\sim$}}}}

\def\beq{\begin{equation}}
\def\eeq{\end{equation}}
\def\be{\begin{equation}}
\def\ee{\end{equation}}
\def\bea{\begin{eqnarray}}
\def\eea{\end{eqnarray}}

\begin{document}

\title{Interpreting the Fermi-LAT gamma ray excess in the simplified framework}

\author{Csaba Bal\'azs$^1$}
\author{Tong Li$^1$}
\author{Chris Savage$^{2,3}$}
\author{Martin White$^4$}

\affiliation{$^1$
ARC Centre of Excellence for Particle Physics at the Tera-scale, School of Physics and Astronomy, Monash University, Melbourne, Victoria 3800, Australia}
\affiliation{$^2$ Nordita, KTH Royal Institute of Technology and Stockholm University, Stockholm, Sweden}
\affiliation{$^3$ Department of Physics \& Astronomy, University of Utah, Salt Lake City, Utah, USA}
\affiliation{$^4$ ARC Center of Excellence for Particle Physics at the Terascale \& CSSM, School of Chemistry and Physics, University of Adelaide, Adelaide, Australia}

\begin{abstract}
We test the plausibility that a Majorana fermion dark matter candidate with a scalar mediator explains the gamma ray excess from the Galactic center.  Assuming that the mediator couples to all third generation fermions we calculate observables for dark matter abundance and scattering on nuclei, gamma, positron, and anti-proton cosmic ray fluxes, radio emission from dark matter annihilation, and the effect of dark matter annihilations on the CMB.
After discarding the controversial radio observation the rest of the data prefers a dark matter (mediator) mass in the 10--100 (3--1000) GeV region and weakly correlated couplings to bottom quarks and tau leptons with values of $10^{-3}$--1 at the $68\%$ credibility level.
\end{abstract}

\maketitle

\section{Introduction}

Since 2009 an increasingly significant deviation from background expectations has been identified in the data of the Large Area Telescope (LAT) on board the Fermi Gamma Ray Space Telescope satellite~\cite{Goodenough:2009gk, Hooper:2010mq, Boyarsky:2010dr, Abazajian:2012pn, Hooper:2013rwa, Gordon:2013vta, Daylan:2014rsa,Calore:2014xka,Murgia}.  The deviation appears around 2 GeV in the energy spectrum of gamma ray flux originating from an extended region centered in the Galactic Center.  The source of the excess photons is unknown.  Their origin can be dark matter (DM) annihilation, a population of millisecond pulsars or supernova remnants~\cite{Carlson:2014cwa, O'Leary:2015gfa, Petrovic:2014xra,Yuan:2014yda}, or cosmic rays injected in a burst-like or continuous event at the galactic center~\cite{Petrovic:2014uda}.  It is, however, challenging to explain the excess with millisecond pulsars~\cite{Cholis:2014lta, Cholis:2014noa} based on their luminosity function.

Recently, several groups including Daylan et al.~\cite{Daylan:2014rsa}, Calore et al.~\cite{Calore:2014xka}, and the Fermi Collaboration~\cite{Murgia} re-analyzed data from the Fermi-LAT~\cite{Atwood:2009ez} and concluded that the 1--3 GeV gamma ray signal is statistically significant and appears to originate from dark matter particles annihilating rather than standard astrophysical sources.  The peak in the energy distribution is broadly consistent with gamma rays originating from self-annihilation of dark matter particles~\cite{Daylan:2014rsa, Gomez-Vargas:2013bea, Abazajian:2014fta, Ipek:2014gua, Ko:2014gha, Cline:2014dwa, Ko:2014loa}.  The intensity of the signal suggests a dark matter annihilation cross section at thermal freeze out ~\cite{Kong:2014haa, Boehm:2014bia, Ghosh:2014pwa, AMartin, Wang:2014elb, Fields:2014pia}.  The diffuse nature and morphology of the gamma ray excess is consistent with a Navarro-Frenk-White-like Galactic distribution of dark matter~\cite{Calore:2014xka}.
This gamma ray excess thus drew the attention of a number of particle model builders and phenomenologists~\cite{Berlin:2014tja,Carlson:2014cwa,Petrovic:2014uda,AMartin,excesspapers,excesspapers15}.

The conclusion that we have discovered dark matter particles, however, cannot be drawn yet.  First, we have to be able to exclude the possibility of a standard astrophysical explanation.  Second, we need to demonstrate that a dark matter particle that explains the gamma ray excess (with a given mass, spin, and interaction strength to the standard sector) is consistent with a large number of other observations.  The latter concerns our paper.  We aim to determine the microscopic properties of the dark matter particle from the gamma ray excess and check that these properties comply with limits from other experiments.  We use dark matter abundance and direct detection data, measurements of the gamma ray flux from the Galactic Center, near Earth positron and anti-proton flux data, Cosmic Microwave Background (CMB) observations, and measurements of galactic radio emission as experimental constraints.

Amongst the above listed experimental bounds the constraining role of radio emission has been debated in the literature.  Bringmann et~al.\ have shown that radio emission imposes severe constraints on dark matter annihilation in the Galactic Center~\cite{Bringmann:2014lpa}.  Radio emission, however, could be induced by various processes including synchrotron radiation, inverse Compton scattering, ionization, and bremsstrahlung.  Most studies of the radio constraint on dark matter, including that of Bringmann et~al., ignore energy loss processes other than synchrotron radiation.  However, as pointed out by Cholis et~al.\ in Ref.~\cite{Cholis:2014fja}, there are several reasons why the other processes could be important.
Cholis et~al.\ have shown that after considering inverse Compton scattering induced by high densities of radiation in the inner Milky Way the radio constraint on dark matter is weakened by about three orders of magnitude~\cite{Cholis:2014fja}.  As a result dark matter annihilating at the thermal rate remains compatible with the radio data.  After considering the effect of diffusion the constraint will be further weakened.  Due to this, we will exclude the radio data point from our combined fit.

As theoretical description of dark matter we use the simplified model framework. Within this ansatz we make minimal and general theoretical assumptions. We consider a single dark matter particle that couples to various standard fermions via a mediator. Our dark matter particle thus annihilates to several final states which all contribute to the observables mentioned above.

This paper is organized as follows. In Sec.~\ref{sec:Models} we introduce the simplified dark matter model we use.
In Sec.~\ref{sec:Constraints}, we describe the observables of dark matter abundance and scattering on nuclei, gamma, $e^+$, and $\bar{p}$ cosmic ray fluxes, and the effect of dark matter annihilations on the CMB.
Our numerical results are given in Sec.~\ref{sec:Results}.
Finally in Sec.~\ref{sec:Concl} we summarize our main results.
We collect the formulae of Bayesian inference and likelihood functions in the Appendix.
\section{Theoretical hypothesis}
\label{sec:Models}

In this section we motivate and describe the theoretical hypothesis we test.  In Ref.~\cite{Balazs:2014jla} we compared Bayesian evidences for three leading simplified models to explain the gamma ray excess from the Galactic center.  We found that the experimental data, especially dark matter direct detection, clearly preferred a Majorana fermion dark matter particle coupled to Standard Model (SM) fermions via a real scalar.
Motivated by this we assume that the dark matter particle is a Majorana fermion, which we denote with $\chi$.  Inspired by the Higgs portal mechanism \cite{Higgsportal}, we use a simplified model to describe interactions between $\chi$ and SM matter.  We assume that the dark-standard mediator is a real scalar field, $S$, and the form of the dark matter to mediator coupling is
\begin{eqnarray}
{\cal L}_\chi \supset \frac{i\lambda_\chi}{2}\bar{\chi}\gamma_5\chi S.
\label{eq:interaction0}
\end{eqnarray}
The presence of $\gamma_5$ is essential since it is lifting the velocity suppression that one otherwise encounters in the indirect detection cross section, thus making this operator capable of explaining the gamma ray excess.
The interaction between the mediator and SM fermions $f$ is assumed to be
\begin{eqnarray}
{\cal L}_S \supset \lambda_f \bar{f}f S.
\label{eq:intercation1}
\end{eqnarray}
In line with minimal flavor violation~\cite{D'Ambrosio:2002ex}, we only consider the third generation fermions, i.e. $f=b,t,\tau$.

For simplicity we assume that mediator pair final states are not present in the dark matter annihilation and only consider s-channel annihilation diagrams.  According to power counting of the dark matter transfer momentum or velocity~\cite{Kumar:2013iva}, with the bi-linears in Eqs.~(\ref{eq:interaction0}) and (\ref{eq:intercation1}) the annihilation cross section of the fermionic dark matter candidate is not velocity suppressed, that is $\sigma v\sim 1$.  The dark matter-nucleon elastic scattering cross section is spin-independent (SI) and momentum suppressed.

\section{Observables}
\label{sec:Constraints}

In this section we describe the calculation of the observables that we use to constrain the parameter space of our hypothesis.  TABLE \ref{tab:Observables} summarizes these observables.

\def\arraystretch{1.4}
\begin{table}[h]
    \begin{tabular}{|c|c|c|c|c|}
        \hline
        observable name & expression & experiment &  data points & data source \\
        \hline
        dark matter abundance        &                                      $\Omega_{\rm DM} h^2$ &       PLANCK & 1 & Ref.\cite{Ade:2013zuv} \\
        \hline
        $\gamma$-ray flux            & $\displaystyle \frac{d^2\Phi_\gamma}{dE d\Omega}$ &    Fermi-LAT & 24 & Ref.~\cite{Calore:2014xka} \\
        \hline
        cosmic $e^+$-ray flux      &              $\displaystyle \frac{d\Phi_{e^+}}{dE}$ &       AMS-02 & 72 & Ref.~\cite{AMS-positron} \\
        \hline
        cosmic $\bar{p}$-ray flux    &        $\displaystyle \frac{d\Phi_{\bar{p}}}{dE}$ &       PAMELA & 23 & Ref.\cite{Adriani:2012paa} \\
        \hline
        Cosmic Microwave Background  &                       $\displaystyle f_{\rm eff}$ &       PLANCK & 3 & Ref.~\cite{Cline:2013fm} \\
        \hline
          dark matter direct detection &                   $\displaystyle s$ &          LUX &   1 & Ref.~\cite{LUX} \\
        \hline
        radio emission               &                             $\displaystyle S_\nu$ & Jodrell Bank &   1 & Ref.~\cite{radio} \\
        \hline
    \end{tabular}
    \caption{Summary of observables we use to constrain our dark matter scenario.  The expressions in the second column are defined in the text of this section.}
    \label{tab:Observables}
\end{table}
\def\arraystretch{1.0}

\subsection{Dark matter abundance}

We assume that dark matter particles, as standard thermal relics, have frozen out in the early universe acquiring their present abundance.  We calculate this abundance using micrOmegas version 3.6.9 \cite{MO}.  We imagine that $\chi$ is the only dark matter candidate, that is we use a Gaussian likelihood function with a mean and width determined by PLANCK~\cite{Ade:2013zuv}
\begin{eqnarray}
\Omega_{\rm DM} h^2 = 0.1199\pm 0.0027.
\label{omega}
\end{eqnarray}
It is challenging to estimate the theoretical uncertainty of the abundance calculation in a simplified model and the task is the subject of a separate paper.  In supersymmetric models, for example, the theoretical uncertainty is comparable to the experimental one over the bulk of the parameter space.  Based on this, we assume an extra theoretical uncertainty of the same size as the experimental error.

\subsection{Gamma ray flux from the Galactic center}

In the theoretical scenario under scrutiny the excess gamma ray flux observed by Fermi-LAT is generated by the self-annihilation of $\chi$ particles.  The differential flux of photons as the function of energy $E$ and observation region $\Omega$ is given by
\begin{eqnarray}
\frac{d^2\Phi_\gamma}{dE d\Omega}=\frac{\langle \sigma v\rangle}{8\pi m_\chi^2}J(\psi)\sum_f B_f \frac{dN_\gamma^f}{dE}.
\label{flux}
\end{eqnarray}
Here $\langle \sigma v\rangle$ is the velocity averaged dark matter annihilation cross section at the Galactic center, $B_f=\langle \sigma v\rangle_f/ \langle \sigma v\rangle$ is the annihilation fraction into the $f{\bar f}$ final state,
and $dN_\gamma^f/dE$ is the energy distribution of photons produced in the annihilation channel with final state $f{\bar f}$.
The $J$ factor in Eq.~(\ref{flux}) is a function of the direction of observation $\psi$
\begin{eqnarray}
J(\psi)=\int_{los}\rho_\chi^2(r)dl,
\end{eqnarray}
with
\begin{eqnarray}
r=\sqrt{l^2+r_\odot^2-2lr_\odot \cos\psi} .
\end{eqnarray}
The dark matter distribution in the Galaxy is described by a generalized Navarro-Frenk-White (NFW) dark matter profile \cite{NFW}
\begin{eqnarray}
\rho_\chi(r)=\rho_0\frac{(r/r_s)^{-\gamma}}{(1+r/r_s)^{3-\gamma}}.
\end{eqnarray}
Here $r_s=20$ kpc is the radius of the galactic diffusion disk, $r_\odot=8.5$ kpc is the solar distance from the Galactic center, and $\rho_0$ is set to reproduce the local dark matter density $\rho_\chi(r_\odot)=0.3 \ {\rm GeV/cm^3}$.
Following Refs.~\cite{Daylan:2014rsa,Calore:2014xka} we fix the inner slope of the NFW halo profile to $\gamma=1.26$ and set $\psi=5^\circ$ in order to avoid bremsstrahlung and other secondary processes~\cite{AMartin}.

The differential yield $dN_\gamma^f/dE$ is different for the three final states we consider.  As seen from Eq.~(\ref{flux}), the total differential yield determining the gamma ray flux is the annihilation-fraction-weighted sum of the differential yields into specific final states. We sum over the contributions of the three individual SM fermions ($b,t,\tau$).  As $B_f$ depends on the model parameters, the gamma ray data plays an important role in constraining the coupling of the mediator to SM fermions.

We use micrOmegas version 3.6.9 to evaluate the theoretical prediction for the differential gamma ray flux~\cite{MO}.  The gamma ray spectral data points that we input into our Gaussian likelihood function are taken from Ref.~\cite{Calore:2014xka}, including both statistical errors and empirical model systematics.

\subsection{Cosmic positron flux near Earth}

The third generation fermion states produced by dark matter annihilation in our model can produce stable leptons in a variety of ways, including production via the decay of top quarks or tau leptons, or secondary production from hadron decays.
These charged particles provide extra sources of cosmic flux in addition to the expected astrophysical backgrounds.
Consequently the measurement of the electron and positron flux allows us to set constraints on the dark matter properties.
The predicted extra electron and positron flux from dark matter annihilation is comparable to that of the secondary production of electrons and positrons, which is one order of magnitude smaller than the measured electron flux itself.
Thus dark matter annihilation affects more the positron flux than the electron flux.
Since the prediction of the electron flux poses an additional challenge and it is the source of considerable uncertainties, we only focus on the positron flux and do not consider the electron flux or the positron to electron fraction in this paper.

The propagation of positrons within the Galaxy is well-described by the following simplified transport equation
\begin{eqnarray}
\frac{\partial f_{e^+}}{\partial t}-\nabla (K(E,r)\nabla f_{e^+})-\frac{\partial}{\partial E}(b(E,r)f_{e^+})=Q_{e^+}(E,r) ,
\label{eq:e+}
\end{eqnarray}
in the diffusion zone approximated by a cylinder with thickness $2L$.
In the above equation $f_{e^+}(r,t,E)$ is the number density of positrons, $K(E,r)$ is the diffusion coefficient which is parameterized as $K(E,r)=K_0(E/{\rm GeV})^\delta$, and $b(E,r)$ is the rate of energy loss.  The source term reads as
\begin{eqnarray}
Q_{e^+}(E,r)=\frac{\rho_\chi^2(r)\langle \sigma v\rangle}{2 m_\chi^2}\left(\sum_f B_f \frac{dN_{e^+}^f}{dE}\right) ,
\end{eqnarray}
with $dN_{e^+}^f/ dE$ being the energy spectrum of positrons produced in the annihilation channel into $f\bar{f}$.  The differential positron flux is given by
\begin{eqnarray}
\frac{d\Phi_{e^+}}{dE}=\frac{v_{e^+}}{4\pi}f_{e^+} ,
\end{eqnarray}
with $v_{e^+}$ being the positron velocity.

For the dark matter induced positron flux calculation in micrOmegas, we take the MED model for the above diffusion parameters: the index of the diffusion coefficient $\delta=0.7$, the normalization factor $K_0=0.0112 \ {\rm kpc^2/Myr}$, and the thickness of the diffusive cylinder $L=4 \ {\rm kpc}$~\cite{Belanger:2010gh}. For the astrophysical backgrounds, we adopt the following parametrization for the interstellar positron flux and the flux at the top of the atmosphere (TOA)~\cite{Ibarra:2013zia}
\begin{eqnarray}
\Phi^{\rm bkg}_{e^+}(E)=C_{e^+}E^{-\gamma_{e^+}}+C_sE^{-\gamma_s}{\rm exp}(-E/E_s) , \\
\Phi^{\rm TOA}_{e^+}(E)=\frac{E^2}{(E+\phi_{e^+})^2}\Phi^{\rm bkg}_{e^+}(E+\phi_{e^+}) ,
\end{eqnarray}
with best-fit parameters $C_{e^+}=72 \ {\rm s^{-1} \ sr^{-1} \ m^{-2} \ GeV^{-1}}$, $\gamma_{e^+}=3.7$, $C_s=1.6 \ {\rm s^{-1} \ sr^{-1} \ m^{-2} \ GeV^{-1}}$, $\gamma_s=2.51$, $E_s=1 \ {\rm TeV}$, and solar modulation parameter $\phi_{e^+}=0.93 \ {\rm GV}$ obtained in Ref.~\cite{Ibarra:2013zia}.
In the above parametrization the potential obtains two contributions, one from the collisions of cosmic rays in the interstellar medium and another from the interactions of high-energy photons.

As experimental input for the positron flux we use the new release of AMS-02 data \cite{AMS-positron}.  We assume the theoretical uncertainty is the same as the AMS-02 experimental error and the form of the likelihood is a composite Gaussian~\cite{Jin:2014ica, Jin:2015sqa}.


\subsection{Cosmic anti-proton flux near Earth}

The propagation of anti-protons through the Galactic cylinder follows a similar diffusion equation as Eq.~(\ref{eq:e+}) but there is an additional effect from the galactic wind and the source term includes the annihilation of anti-protons in the interstellar medium as well as the annihilation of dark matter. The energy loss of anti-protons, however, is negligible compared with that of the positrons.

The astrophysical background is calculated by adopting the set of propagation parameters called the KRA model in Ref.~\cite{Cirelli:2014lwa}.  These parameters were extracted from a fit to standard cosmic ray data.
To calculate the anti-proton flux from dark matter annihilation, i.e. $d\Phi_{\bar{p}}/dE$, we use the same MED model described in the last subsection and assume the velocity of the convective wind to be $V_{\rm conv}=12 \ {\rm km/s}$. The anti-proton flux at low energies is also altered by solar modulation effects.  We use the Fisk potential $\phi_F$, which relates the local interstellar anti-proton flux to the one measured at the top of the atmosphere, as described in the KRA model.

We use the latest release of PAMELA data as experimental input for the anti-proton flux~\cite{Adriani:2012paa}.  Note that the error bars in this data release are only statistical. Systematic error bars are expected to be of the same order of magnitude as in the first release of PAMELA data~\cite{Adriani:2010rc}.  We combine the uncertainties in quadrature and assume that the theoretical uncertainty is the same as the experimental error in the composite Gaussian likelihood function.


\subsection{Cosmic Microwave Background}


Dark matter annihilation in the early universe affects the CMB temperature and polarization fluctuations.  Thus the CMB power spectrum measurement from PLANCK provides constraints on dark matter properties.  A key quantity for determining the constraint on a given dark matter model is the efficiency for producing ionizing radiation.  The authors of Ref.~\cite{Cline:2013fm} provide values of the effective efficiency $f_{\rm eff}$ for different annihilation channels and dark matter masses that can be easily interpolated.  We quantify the CMB constraints using the following likelihood function
\begin{eqnarray}
\mathcal{L}_{\rm CMB}&=&{\rm exp}\left[-\frac{1}{2}f_{\rm eff}^2\lambda_1c_1^2\left(\frac{\langle\sigma v\rangle}{2\times 10^{-27}{\rm cm}^3{\rm s}^{-1}}\right)^2\left(\frac{{\rm GeV}}{m_\chi}\right)^2\right],\\
f_{\rm eff}&=&\sum_{i=\tau,b,t}f_{{\rm eff},i}B_i,
\label{eq:CMBL}
\end{eqnarray}
with $\lambda_1=3.16$ and $c_1=4.64$ for the PLANCK data.
Here $B_i$ is the annihilation branching fraction defined earlier.



\subsection{Dark matter direct detection}

Direct detection of dark matter is facilitated by dark matter particles scattering on nuclei of a target material in a well shielded detector.  The differential recoil rate of dark matter on nuclei, as a function of the recoil energy $E_R$, is
\begin{eqnarray}
\frac{dR}{dE_R}=\frac{\rho_\chi}{m_\chi m_A}\int dv v f(v) \frac{d\sigma_A^{\rm SI}}{dE_R} ,
\end{eqnarray}
where $m_A$ is the nucleus mass, $f(v)$ is the dark matter velocity distribution function and
\begin{eqnarray} \label{dsigmadE}
\frac{d\sigma_A^{\rm SI}}{dE_R} = G^\chi(q^2) \frac{1}{E_{\rm max}} \frac{4\mu^2_A}{\pi} [Zf_p^\chi+(A-Z)f_n^\chi]^2 F_A^2(q),
\end{eqnarray}
with $E_{\rm max}=2\mu^2_A v^2/m_A$, 
$G^{\chi}(q^2)=\frac{q^2}{4m_\chi^2}$~\cite{Berlin:2014tja},
and $f_N^{\chi}=\frac{\lambda_\chi}{2m_S^2}g_{SNN} \ (N=p,n)$. 
$F_A(q)$ is the nucleus form factor and $\mu_A=m_\chi m_A/(m_\chi+m_A)$ is the reduced dark matter-nucleon mass.
We assume that the local disk rotation speed is $220 \ {\rm km/s}$ with the same value for the most probable speed of the dark matter's Maxwell-Boltzmann velocity distribution. The Galactic escape speed is $544 \ {\rm km/s}$~\cite{LUX}.

As we only consider the interaction mediated by the scalar between the dark matter particles and the third generation quarks, the strength of the mediator-nucleon ($N$) interaction reads
\begin{eqnarray}
g_{SNN}=\frac{2}{27}m_Nf_{TG}\sum_{f=b,t}\frac{\lambda_f}{m_f}.
\end{eqnarray}
Above $f_{TG}=1-f_{T_u}^N-f_{T_d}^N-f_{T_s}$ and we adopt $f_{T_u}^p=f_{T_d}^n=0.02$, $f_{T_d}^p=f_{T_u}^n=0.026$, $f_{T_s}=0.043$~\cite{Berlin:2014tja,lattice,Crivellin:2013ipa}.

For the LUX likelihood function, we use a Poisson distribution in the observed number of events $N$,
\begin{equation} \label{Poisson}
\mathcal{L}(s|N)
= P(N|s)
= \frac{(b+s)^N \, e^{-(b+s)}}{N!} \, ,
\end{equation}
where $b$ is the expected number of background events,
\begin{equation} \label{signal}
s = MT \int_0^{\infty} dE \; \phi(E) \, \frac{dR}{dE_R}(E)
\end{equation}
is the expected signal, $MT$ is the detector mass$\times$time exposure, and $\phi(E)$ is a global efficiency factor that takes into account trigger efficiencies, energy resolution, and analysis cuts.  Likelihood calculations are performed using a version of \texttt{LUXCalc} \cite{Savage:2015xta} modified to include the additional momentum dependence in Eqn.~(\ref{dsigmadE}).  For the LUX analysis region used by \texttt{LUXCalc}, $N=1$ and $b=0.64$; the efficiency curve $\phi(E)$ was generated by \texttt{TPCMC} \cite{Savage:2015tpcmc} using the \texttt{NEST} model \cite{Szydagis:2011tk,Szydagis:2013sih}.  See Ref.~\cite{Savage:2015xta} for further details.



\subsection{Radio Emission}

Electrons and positrons from dark matter annihilation are expected to lose energy through synchrotron radiation in the presence of large scale magnetic fields.  Thus the radio emission in galaxies and galaxy clusters can also be used to place constraints on the dark matter properties.  The synchrotron flux density is given by
\begin{eqnarray}
S_\nu = \frac{1}{4\pi} \frac{J}{\rho_\chi^2} \int 2 \frac{d\Phi_{e^+}}{dE}\frac{dW_{\rm syn}}{d\nu} dE_e ,
\end{eqnarray}
where $d\Phi_{e^+}/dE$ is the positron flux in units of $\rm (GeV \ cm^2 \ s \ sr)^{-1}$. The synchrotron power per frequency reads
\begin{eqnarray}
\frac{dW_{\rm syn}}{d\nu}=\frac{\sqrt{3}}{6\pi}\frac{e^3B}{m_e}F\left(\frac{\nu}{\nu_{\rm syn}}\right) ,
\end{eqnarray}
with
\begin{eqnarray}
F(x)=x\int_x^\infty K_{5/3}(\xi)d\xi\approx \frac{8\pi}{9\sqrt{3}}\delta(x-1/3) .
\end{eqnarray}
The $\delta$-function implies
\begin{eqnarray}
\nu_{\rm syn}=3\nu=\frac{3eBp^2}{4\pi m_e^3}\approx \frac{3eBE_e^2}{4\pi m_e^3} .
\end{eqnarray}

For simplicity, we fix the magnetic field strength at a conservative lower limit \cite{Bringmann:2014lpa}
\begin{eqnarray}
B=50 \ \mu {\rm G}.
\end{eqnarray}
The integration cone in the $J$ factor corresponds to a $4''$ region around the Galactic center.

\section{Results}
\label{sec:Results}

We coded the Lagrangian of the relevant simplified dark matter model in FeynRules~\cite{FR}.  Calculation of observables, including the dark matter relic density and nucleon scattering interactions, differential gamma ray, $e^+$ and $\bar{p}$ fluxes, and radio signal were performed using a modified version of micrOmegas 3.6.9~\cite{MO}.  Nested sampling and posterior distribution calculations were performed by MultiNest~\cite{MN}.  The nested sampling algorithm was developed to calculate marginalized posterior probability distributions and it is a Bayesian's way to numerically implement Lebesgue integration \cite{Skilling2006}.
Since the relevant part of the likelihood distribution spans multiple orders of magnitude, we use log priors for all parameters. We present further details of our statistical analysis in the Appendix.

\begin{table}[h]
    \begin{tabular}{|c|c|c|c|c|c|}
        \hline
        parameter & $m_\chi$ & $m_S$  &  $\lambda_b$ & $\lambda_t$ & $\lambda_\tau$ \\
        (unit)    &    (GeV) &  (GeV) &              &             &                \\
        \hline
        scan range & $1-10^3$ & $1-10^3$ & $10^{-5}-10$ & $10^{-5}-10$ & $10^{-5}-10$ \\
        \hline
        prior type & log & log & log & log & log \\
        \hline
    \end{tabular}
    \caption{Scan ranges and prior types used for the scanned parameters.}
    \label{tab:ScanRange}
\end{table}

In our numerical calculation we fix the dark matter to mediator coupling as $\lambda_\chi=1$, and we scan the following free parameters:
\begin{eqnarray}
 P = \{ m_\chi, m_S, \lambda_b, \lambda_t, \lambda_\tau \}.
\label{eq:para}
\end{eqnarray}
Here $m_\chi$ is the mass of the dark matter particle, $m_S$ is the mass of the scalar mediator, and $\lambda_f \ (f=\tau,b,t)$ is the coupling of the mediator to the SM fermion pair $f{\bar f}$ as defined in Eqs.~(\ref{eq:interaction0}) and~(\ref{eq:intercation1}).  The range of our scan over the above parameters and the type of prior we use is given in TABLE~\ref{tab:ScanRange}.

\begin{figure}[t]
\begin{center}
\includegraphics[scale=1,width=8cm]{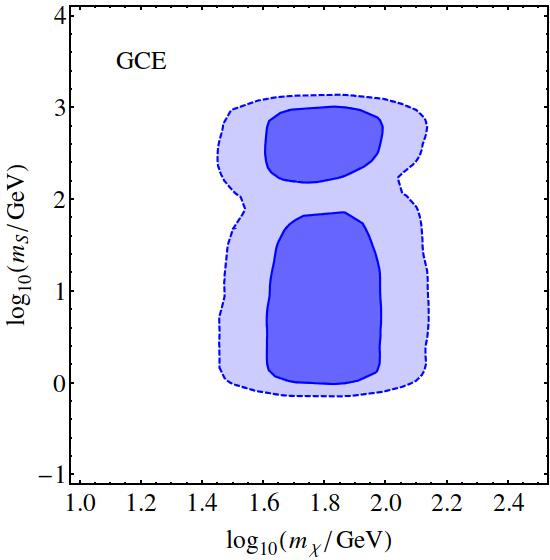}
\includegraphics[scale=1,width=8cm]{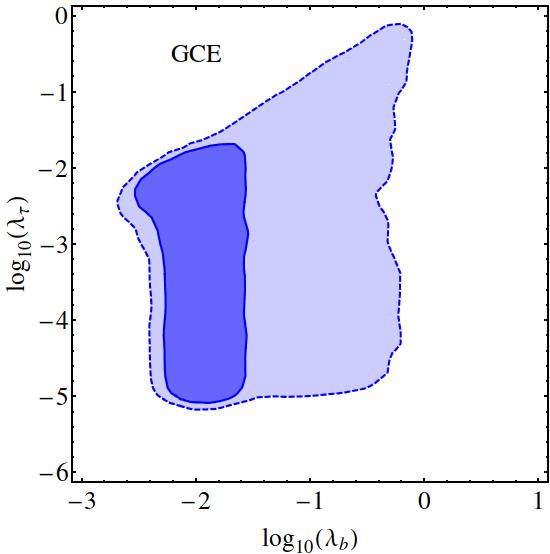}
\end{center}
\caption{Posterior probability distributions marginalized to the scanned model parameters.  The likelihood function for these plots only contains the dark matter abundance and the anomalous Fermi-LAT gamma ray data. The dark and light regions hereinafter correspond to $68\%$ and $95\%$ credible regions, respectively.}
\label{fig:gammaRay}
\end{figure}

To build some intuition, first we examine the constraining effect of each observable one by one.  To this end we plot the posterior probability distributions marginalized to the scanned model parameters such that the likelihood function only contains the dark matter abundance and one of the other observables.
In Fig.~\ref{fig:gammaRay} we show marginalized posterior probability distributions taking into account the dark matter abundance and the gamma ray data.  The first frame of Fig.~\ref{fig:gammaRay} confirms that the gamma ray data restrict the range of the dark matter mass close to 35--60 GeV~\cite{Daylan:2014rsa,Calore:2014xka}.  It is less appreciated, however, that uncertainties still allow a 40--100 (25--160) GeV dark matter mass range at the $68\%$ ($95\%$) credibility level.  The gamma ray data, coupled with the dark matter relic density, allows the whole mass range of the scalar mediator.
It is also interesting to note that the preferred dark matter mass region is dissected by a diagonal band with a lower posterior around the on-shell resonance region $m_S = 2 m_\chi$.  In this valley, dark matter resonantly annihilates via the $s$-channel mediator, depleting its abundance.  Thus, it is harder for the model to match the PLANCK constraint.

The relevant interaction strengths also remain virtually unconstrained as shown by the right frame of Fig.~\ref{fig:gammaRay}.  PLANCK and the anomalous Fermi-LAT gamma ray data only restrict these coupling in the $\lambda_\tau=1\times10^{-5}$ -- $2.5\times10^{-2}$  ($6\times10^{-6}$ -- 1) and $\lambda_b=2.5\times10^{-3}$ -- $2.5\times10^{-2}$  ($1.6\times10^{-3}$ -- 1) ranges at the $68\%$ ($95\%$) credibility level.  Simultaneous order 1 couplings are marginally allowed and appear in the part of the parameter space where the annihilation cross section is suppressed by sizable $m_S$.

\begin{figure}[t]
\begin{center}
\includegraphics[scale=1,width=8cm]{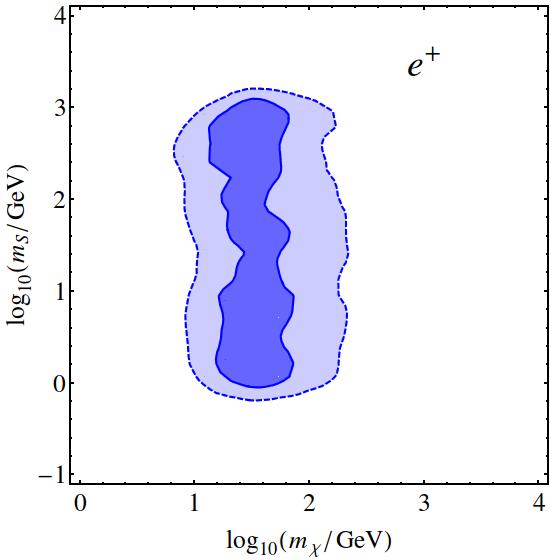}
\includegraphics[scale=1,width=8cm]{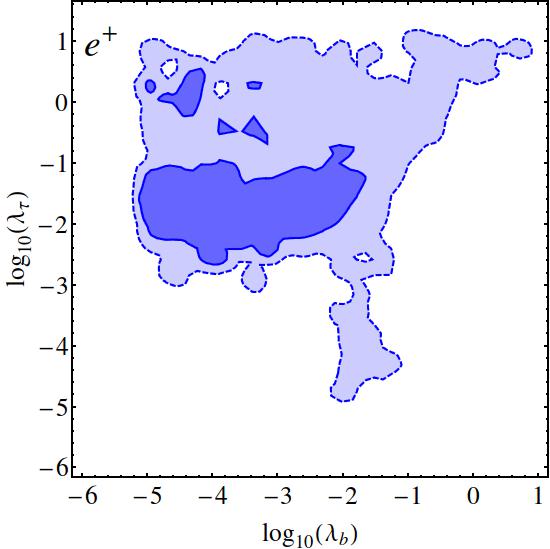}
\end{center}
\caption{Posterior probability distributions marginalized to the scanned model parameters.  The likelihood function for these plots only contains the dark matter abundance and the AMS-02 positron flux data.}
\label{e+}
\end{figure}

In Fig.~\ref{e+} we show marginalized posterior probability distributions with the likelihood function containing only the dark matter abundance and the AMS-02 positron flux data.  The AMS-02 measurement of the positron flux features a small upward kink, a sudden change of slope, around 35 GeV.  A smooth background prediction has a hard time to reproduce this kink and systematically falls below the experimental points in the 35--50 GeV region.  Positrons originating from the annihilation of a 35-50 GeV dark matter particle can fill the gap between the background and the data.  Hence the AMS-02 data show a mild preference toward a dark matter candidate with 16--65 (10--160) GeV mass at $68\%$ ($95\%$) credibility level.  AMS-02 also restricts the dominant decay to $\tau$ leptons with a $\lambda_\tau$ coupling around $2.5\times 10^{-3}-0.1$ and $\lambda_b$ coupling below about $10^{-2}$ at the $68\%$ credibility level.

\begin{figure}[t]
\begin{center}
\includegraphics[scale=1,width=8cm]{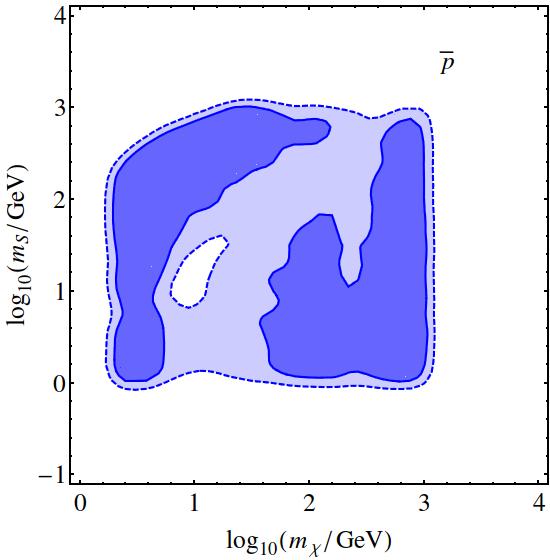}\\
\includegraphics[scale=1,width=8cm]{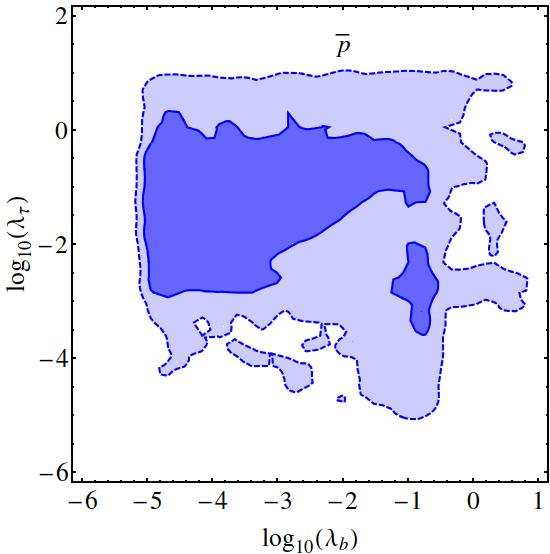}
\includegraphics[scale=1,width=8cm]{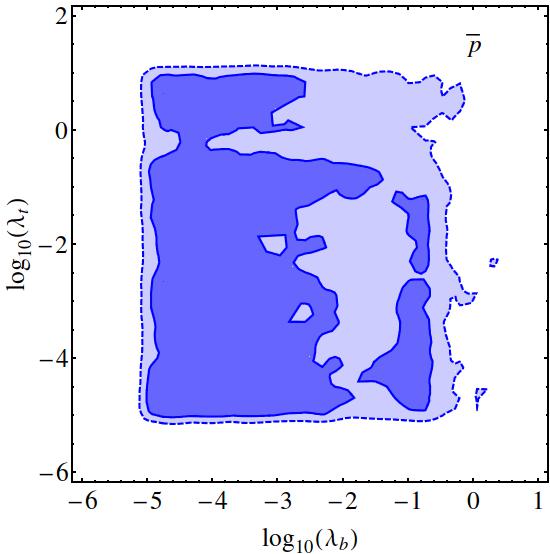}
\end{center}
\caption{Posterior probability distributions marginalized to the scanned model parameters.  The likelihood function for these plots only contains the dark matter abundance and the PAMELA anti-proton cosmic ray flux data.}
\label{pbar}
\end{figure}

Fig.~\ref{pbar} shows marginalized posterior probability distributions with the likelihood function including the PLANCK and PAMELA anti-proton cosmic ray flux data only.  The PAMELA data in itself does not prefer any particular parameter region.  Dark matter and mediator masses are both allowed in the full scanned range at the $95\%$ credibility level.  This happens with the exception of a small island around $m_\chi \sim m_S \sim 10$ GeV where the combined PLANCK and PAMELA constraints are harder to satisfy.  The reasons for this are that this island falls on the $m_S = 2 m_\chi$ resonant annihilation corridor and the PAMELA data around 10 GeV leave very little room for dark matter.  This situation improves for lower dark matter masses.

Since dark matter masses above the top quark mass are allowed by the combination of PLANCK and PAMELA, the $\lambda_t$ coupling comes into play.  These data, however, are not sufficient to constrain $\lambda_t$.  It is interesting to note that PLANCK and PAMELA allow fairly large values of $\lambda_t$, $\lambda_b$, and $\lambda_\tau$ for heavier $m_\chi$ and $m_S$ in order to accommodate the correct relic abundance.

\begin{figure}[t]
\begin{center}
\includegraphics[scale=1,width=8cm]{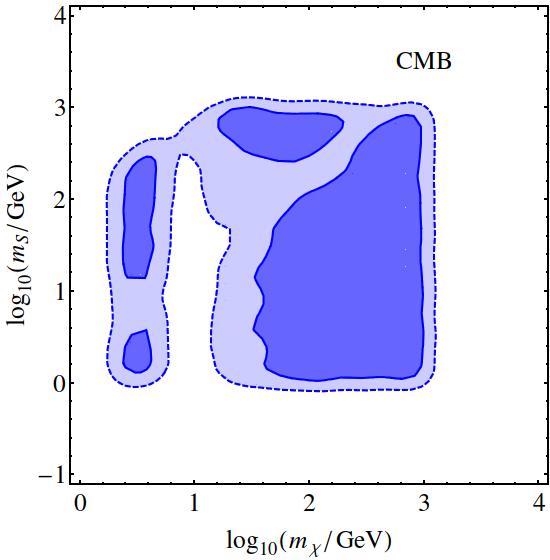}\\
\includegraphics[scale=1,width=8cm]{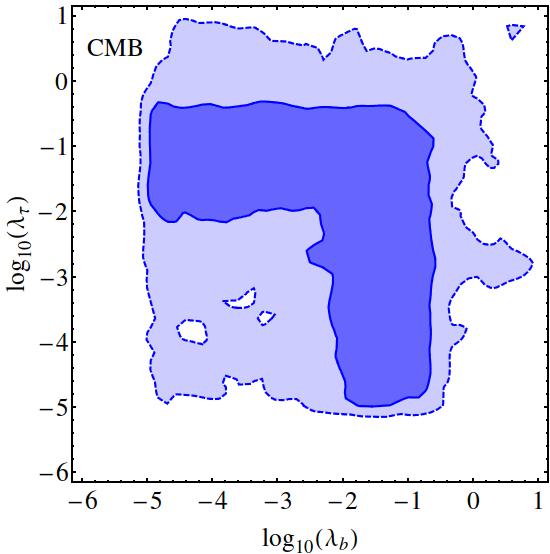}
\includegraphics[scale=1,width=8cm]{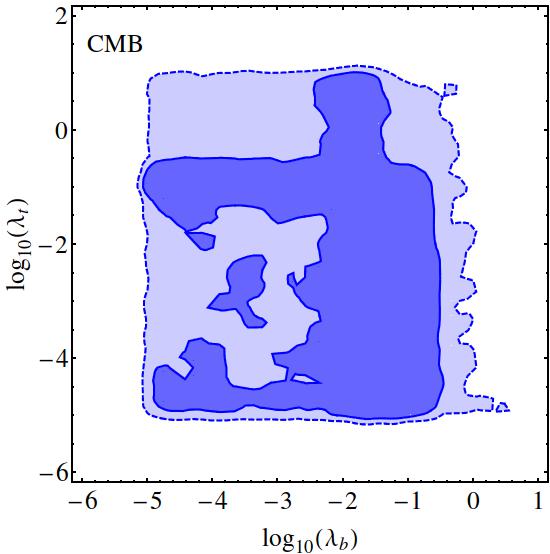}
\end{center}
\caption{Posterior probability distributions marginalized to the scanned model parameters.  The likelihood function for these plots only contains the dark matter abundance and CMB terms, as defined in Eq.~(\ref{eq:CMBL}).}
\label{CMB}
\end{figure}

In Fig.~\ref{CMB} we show marginalized posterior distributions for the dark matter abundance and CMB likelihood function, defined in Eq.~(\ref{eq:CMBL}).  The diagonal depletion of the likelihood function due to dark matter resonant annihilation is apparent in the $m_\chi$ vs $m_S$ frame.  The CMB likelihood function suppresses the posterior around $m_\chi = 10$ GeV providing more constraint on low mass dark matter.

The posterior probability distribution projected to the $\lambda_b$ vs. $\lambda_\tau$ couplings shows a peculiar pattern.  This pattern is the combined result of two relatively simple sets of constraints.  Dark matter abundance is responsible for the low likelihood values at low $\lambda_b$ and $\lambda_\tau$.  It turns out that the PLANCK constraint on the amount of relic dark matter is hard to respect unless one of these couplings is sizable, that is $\lambda_b$ or $\lambda_\tau \gtrsim 10^{-2}$ at the $68\%$ credibility level.  If both of these couplings are small then annihilation is slow and dark matter is overproduced in the early universe.
In the large coupling region, on the other hand, the CMB constrains $\lambda_b$ and $\lambda_\tau$ from above.  If any of these couplings are larger than about 0.1 then dark matter tends to become under-produced and the CMB receives too much modification from dark matter.  The $\lambda_t$ coupling is hardly constrained by the CMB at the $95\%$ credibility level.

\begin{figure}[t]
\begin{center}
\includegraphics[scale=1,width=8cm]{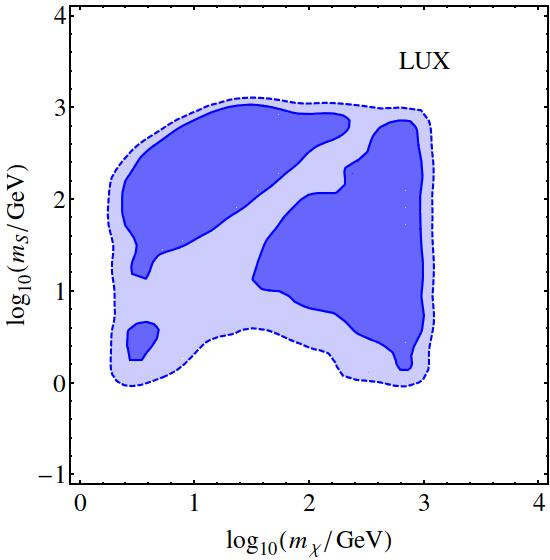}\\
\includegraphics[scale=1,width=8cm]{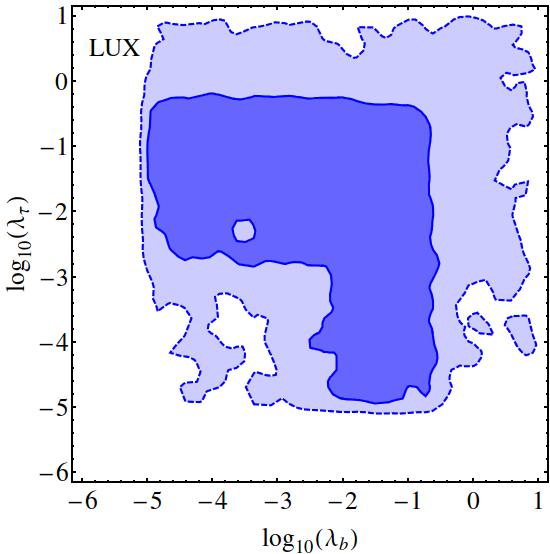}
\includegraphics[scale=1,width=8cm]{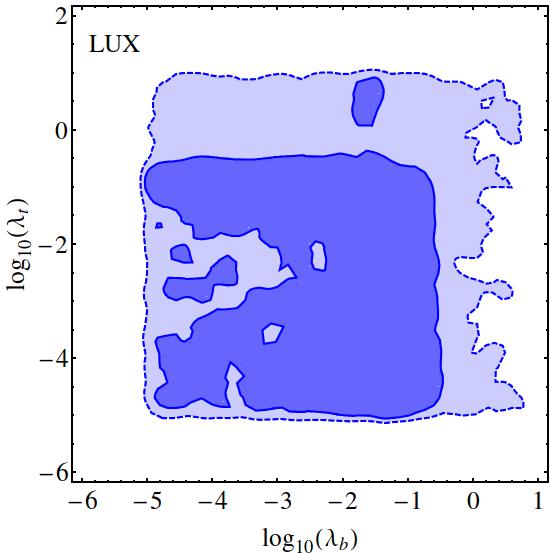}
\end{center}
\caption{Posterior probability distributions marginalized to the scanned model parameters.  The likelihood function for these plots only contains the dark matter abundance and LUX data.}
\label{DD}
\end{figure}

In Fig.~\ref{DD} we show the marginalized posterior distribution with the likelihood function containing only the PLANCK and LUX data.  Due to momentum suppression of the nucleon-$\chi$ elastic scattering, the LUX data very weakly constrain the dark matter or mediator mass.  In the $m_\chi$ vs. $m_S$ plane the diagonal resonant annihilation valley is visible, but no other structure is present.  The posterior probability distribution for the couplings is very similar to that in Fig.~\ref{CMB}.  Similarly to the case of the CMB, PLANCK and LUX only impose a constraint on the order 1 couplings.

As discussed in the Introduction, the radio signal potentially  very strictly constrains dark matter~\cite{Bringmann:2014lpa}.  Assuming that dark matter contributes to the radio signal only by synchrotron radiation we find the radio flux upper limit of Jodrell Bank at 408 MHz~\cite{radio} excludes the dark matter hypothesis we consider by two orders of magnitude.  Our finding fully confirms that of Ref.~\cite{Bringmann:2014lpa}.
This exclusion, on the other hand, is lifted if inverse Compton scattering, ionization, and bremsstrahlung are also considered as potential dark matter energy loss mechanisms leading to radio emission~\cite{Cholis:2014fja}.  As shown by Ref.~\cite{Cholis:2014fja} the bound from the radio data is weakened by about three orders of magnitude if inverse Compton scattering is considered and is expected to pose no constraint after including Galactic diffusion effects.
Due to this, we do not include the radio emission data point in our combined likelihood function.

\begin{figure}[t]
\begin{center}
\includegraphics[scale=1,width=8cm]{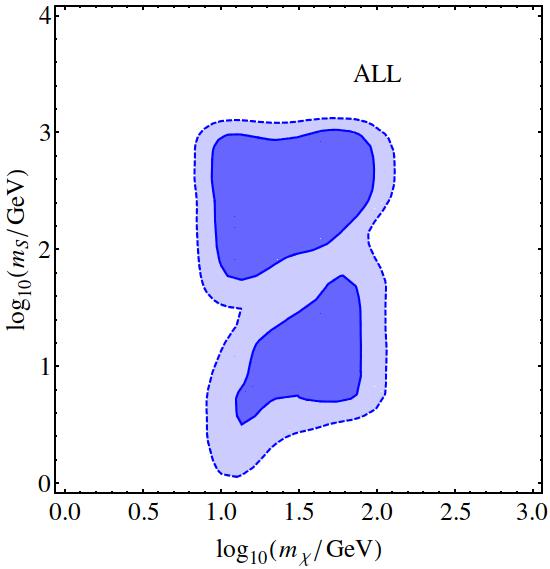}
\includegraphics[scale=1,width=8cm]{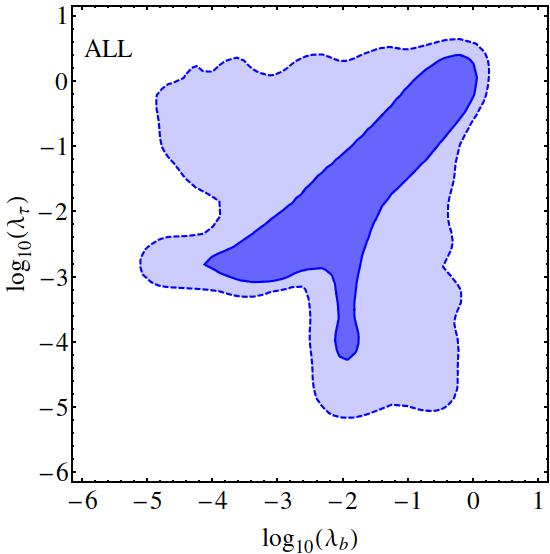}
\end{center}
\caption{Posterior probability distributions marginalized to the scanned model parameters.  The likelihood function for these plots contains all of dark matter abundance, Fermi-LAT gamma ray data, AMS-02 positron flux data, PAMELA anti-proton ray flux data, CMB and LUX data.}
\label{fig:combined}
\end{figure}

The summary of all constraints is presented in Fig.~\ref{fig:combined}.  The constraint on the dark matter mass is dominated by the gamma ray data and the final combination restricts $m_\chi$ to the 10--100 (7--125) GeV region with $68\%$ ($95\%$) credibility.  Less of the low mediator mass region survives the scrutiny of the combined constraints, leaving the 3--1000 GeV $m_S$ region preferred at the $68\%$ credibility level.
The combined constraints prefer a somewhat correlated pair of $\lambda_\tau$ and $\lambda_b$ couplings in the intermediate $10^{-3}$--1 region at the $68\%$ credibility level. Small ($\lambda < 10^{-3}$) and large ($\lambda > 1$) values of couplings are disfavored mostly by PLANCK at the $95\%$ credibility. Comparing the second frame of Fig.~\ref{fig:combined} to those showing the individual constraints it is clear that simultaneous order 1 couplings are mildly under stress from almost all the data.

\section{Conclusions}
\label{sec:Concl}

In this work we perform a comprehensive statistical analysis of the gamma ray excess from the Galactic Center in a simplified dark matter model framework. According to our previous study, Majorana fermion dark matter interacting with standard model fermions via a scalar mediator is the most favoured explanation of the galactic center excess when characterised by Bayesian evidence.
We locate the most plausible parameter regions of this theoretical hypothesis using experimental data on the dark matter abundance and direct detection interactions, the gamma ray flux from the Galactic center, near Earth positron and anti-proton fluxes, the Cosmic Microwave Background, and galactic radio emission.

We find that the radio data excludes the model if we include synchrotron radiation as the only energy loss channel.  Since it was shown that inclusion of other types of energy losses lifts this exclusion we discard the single radio data point from our combined likelihood~\cite{Cholis:2014fja}.
The rest of the data prefers a dark matter (mediator) mass in the 10--100 (3--1000) GeV region and weakly correlated couplings to bottom quarks and tau leptons with values of $10^{-3}$--1 at the $68\%$ credibility level.

\acknowledgments
We thank Pat Scott for helping with the CMB constraint and Alejandro Ibarra for providing positron flux data points in Ref.~\cite{Ibarra:2013zia}.
This work in part was supported by the ARC Centre of Excellence for Particle Physics at the Terascale.
%
%
The use of Monash University Sun Grid, a high-performance computing facility, and the National Computational Infrastructure (NCI), the Southern Hemisphere's fastest supercomputer, is also gratefully acknowledged.  MJW is supported by the Australian Research Council Future Fellowship FT140100244. CS thanks the School of Physics at Monash University, where part of this work was performed, for its hospitality. CS was partially supported by NSF award 10681111.

\appendix

\section{Bayesian Inference}

In this section we summarize the statistical background of our analysis.  Let $P(A|I)$ and $P(B|I)$ denote the plausibility of two non-exclusive propositions, $A$ and $B$, in light of some prior information, $I$.  The probability that both $A$ and $B$ are correct is given by the conditional expression \begin{equation}
 P(AB|I) = P(A|BI)P(B|I) .
\end{equation}
Bayes theorem follows from the symmetry of the conditional probability under the exchange of $A$ and $B$:
\begin{equation}
 P(A|BI) = \frac{P(B|AI)P(A|I)}{P(B|I)} .
\end{equation}
In this context $P(A|I)$ is typically called the prior probability and represents the plausibility of our hypothesis given information prior the observation $B$.  The likelihood function $P(B|AI)$ indicates how accurately the hypothesis can replicate the data.  The posterior probability $P(A|BI)$ quantifies the plausibility of the hypothesis $A$ given the data $B$.  The evidence $P(B|I)$ serves to normalize the posterior.

For theoretical models with a continuous parameter $\theta$ Bayes' theorem can be recast in the form
\begin{equation}
\mathcal{P}(\theta|B,I) = \frac{\mathcal{L}(B|\theta,I)\pi(\theta,I)}{\epsilon(B,I)} .
\label{eqnBayes}
\end{equation}
The posterior distribution can be used to estimate the most likely region of $\theta$.  The evidence is calculated via an integral over the full parameter space
\begin{equation}
\epsilon(B,I) = \int_\theta \mathcal{L}(B|\theta,I)\pi(\theta,I) d\theta .
\end{equation}
For more than one continuous parameters, $\theta_i$, marginalization is performed by integrating the posterior over various parameters in the higher dimensional parameter space
\begin{equation}
\mathcal{P}(\theta_j) = \int \prod_{i\neq j} d\theta_i \mathcal{P}(\theta_i).
\end{equation}


\end{document}